\documentstyle[11pt,aaspp4,epsf,rotate]{article}

\def\psim{\lower.5ex\hbox{$\; \buildrel \propto \over \sim \;$}}
\begin{document}

\title{On Compton Scattering Scenarios for Blazar Flares}

\author{Markus B\"ottcher\altaffilmark{1,2} \& Charles D.
Dermer\altaffilmark{2}}

\altaffiltext{1}{Department of Space Physics and Astronomy, Rice University,
6100 Main Street, Houston, TX  77005-1892}
\altaffiltext{2}{E. O. Hulburt Center for Space Research, Code 7653,
       Naval Research Laboratory, Washington, DC 20375-5352}

\begin{abstract}

The synchrotron reflection scenario recently proposed to explain 
$\gamma$-ray flares observed from blazar jets is studied. Our 
analysis takes into account the angular distribution of the 
beamed radiation, the finite extent of the scattering region, 
and light travel-time effects. We compare energy densities and 
powers for synchrotron, SSC, reflected synchrotron (RSy), and external
Compton (EC) scattering processes. If the width of the scattering 
layer is much larger than $\Gamma R^\prime_B$, where $\Gamma$ and $
R^\prime_B$ denote the bulk Lorentz factor and comoving-frame
radius of the plasma blob, respectively, then the ratio of the 
RSy and synchrotron energy densities $\sim 4 \,
\Gamma^3 n_{\rm BLR} \sigma_{\rm T} R^\prime_B$, where $n_{\rm BLR}$ 
is the mean particle density in the broad line region (BLR).  
Our results imply that Thomson-thick scattering regions of 
narrow extent must be present for the synchrotron reflection 
mechanism to operate effectively. This process seems unlikely 
to cause flares in lineless BL Lac sources, where X-ray
and TeV flares are common and the BLR is thought to be weak or 
absent. We sketch time profiles of flares for various scenarios,  
including a model where the blob is energized by sweeping up surrounding
material.

\end{abstract}


\section{Introduction}

More than 50 blazar-type AGNs have been detected with high  
confidence by EGRET to emit $\gamma$-rays above 100~MeV  
(Matox et al. \markcite{Mattox97}1997). These sources are  
identified with flat-spectrum radio sources classified as
BL-Lac objects or quasars. Many of these objects exhibit 
variability on all wavelengths, with some of the most rapid variability, 
on time scales of hours to days, observed at the highest $\gamma$-ray  
energies (e.g., Bloom et al. \markcite{bloom97}1997; Wagner et al. 
\markcite{wagner95}1995; Mukherjee et al. \markcite{mukerjee97}1997).

The large apparent luminosities in combination with the short
variability time scales provide evidence for the widely accepted
relativistic jet model for AGNs (for recent reviews, see Schlickeiser
\markcite{schlickeiser96}1996 and Hartman et al. \markcite{Hartman97}1997),
according to which the radio--$\gamma$-ray  emission from blazars is
emitted via nonthermal synchrotron radiation and Comptonization of soft
photons by energetic particles in relativistic outflows powered by
accreting supermassive black holes. Soft photons which
are Compton-scattered to produce the $\gamma$-ray emission include 
internal synchrotron photons (e.g., Marscher \& Gear \markcite{mg85}1985, 
Maraschi et al. \markcite{maraschi92}1992, Bloom \& Marscher 
\markcite{bm96}1996) and accretion-disk radiation which enters the 
jet directly (Dermer \& Schlickeiser \markcite{ds93}1993)
and after being scattered by surrounding BLR clouds and 
circumnuclear debris (e.g., Sikora, Begelman \& Rees 
\markcite{sbr94}1994; Blandford \& Levinson \markcite{bl95}1995; 
Dermer, Sturner, \& Schlickeiser \markcite{dss97}1997; Protheroe \& Biermann 
\markcite{pb97}1997).

It has recently been proposed (Ghisellini \& Madau \markcite{gm96}1996;
hereafter GM96) that the beamed synchrotron radiation, after scattering
off a cloud near the jet trajectory and reentering  the jet, can be a
source of copious soft photons and lead to a pronounced flare of very
short duration as the relativistic jet plasma passes through the cloud. 
Wehrle et al. (\markcite{Wehrle98}1998) argue that this mechanism might explain 
the February 1996 $\gamma$-ray flare observed from 3C 279. A detailed analysis 
which
correctly accounts for causality effects and the finite width of the
scattering layer was not performed by GM\markcite{gm96}96, and such a
treatement is required before  blazar flare spectra and light
curves can be modeled.

Here we examine this model in more detail for a simple geometry of 
the BLR in the limiting regime where the nonthermal jet electron 
distribution does not evolve. In \S 2  we describe the model.  
Numerical calculations of the magnetic-field and photon energy
densities and synchrotron and Compton powers are presented in
\S 3. Application to blazar flares is made and the effects of 
blob energization on time profiles of flares are indicated.  We 
summarize in \S 4.

\section{Description of the Model}

The idealized system, illustrated in Figure 1, is assumed to have 
the following qualities: A plasma blob, which in the comoving frame 
is spherical with radius $R'_B \sim  10^{15}$~--~$10^{17}$~cm, 
moves with bulk Lorentz factor $\Gamma \sim 10$ and velocity 
$\beta_{\Gamma} c$ along the $\hat z$ axis which defines the 
jet.  The blob is filled with relativistic, nonthermal electrons 
and pairs which are isotropically distributed in the comoving 
frame with a comoving power-law Lorentz factor distribution 
$n_e (\gamma) \propto \gamma^{-p}$ for $\gamma_1 \le \gamma \le 
\gamma_2$. The nonthermal lepton density in the jet is denoted 
by $n_{\rm e,jet}$, and a randomly oriented magnetic field $B$,
in equipartition with the energy in relativistic leptons,
is present in the blob. An accretion disk is located at $z = 0$
and radiates with a luminosity $L_D \sim 10^{44}$ --
$10^{46}$~erg~s$^{-1}$.  The disk spectrum is approximated by a thermal
blackbody of $T_{\rm disk} \sim 10^5 $~K peaking  at the characteristic
maximum energy of a standard Shakura-Sunyaev disk spectrum (Shakura \&
Sunyaev \markcite{ss73}1973) for a central black hole of
mass $M \sim 10^8 \, M_{\odot}$.

The accretion-disk/jet system is assumed to be surrounded by a 
spherically distributed layer of scattering material which represents
the BLR.  This layer extends from $r_{\rm in}$ to $r_{\rm out}$ and
has a radial Thomson depth $\tau_{\rm T,BLR} \sim 0.01$~--~0.1 and
a uniform electron density $n_e^{\rm BLR}$. In the following, primed 
quantities are measured in the reference frame comoving with the  
relativistic plasma blob, while unprimed quantities are generally 
referred to the stationary accretion-disk frame. For simplicity, we 
assume that the relativistic electron distribution does not evolve 
during the outward-motion of the blob. This assumption leads to an
overprediction of the contribution from the jet synchrotron mirror 
for which the accumulation of radiation along the jet trajectory is 
important. The synchrotron and Compton processes are calculated using 
$\delta$-function approximations for the emission from relativistic 
electrons (see, e. g., Dermer et al.\ \markcite{dss97}1997).

The important invariant with which to compute the energy density in
accretion-disk radiation scattered by the BLR (denoted by the superscript 
`ECC' for EC radiation from Clouds) relates the angle-dependent
photon number density between the comoving and the
stationary frames, given by $n'_{\rm ph} (\epsilon',\Omega')
= [\Gamma(1+\beta_\Gamma\mu')]^{-2}n_{\rm ph} (\epsilon,\Omega)$, 
where $\epsilon = h\nu/(m_e c^2)$ is the dimensionless photon energy. 
Following the treatment outlined by B\"ottcher \& Dermer 
(\markcite{bd95}1995), we find that the comoving-frame
photon number density at height $z_2$ is given by

\begin{equation} {n'}_{\rm ph}^{\rm ECC} (\epsilon'_2; z_2) =
{n_e^{\rm BLR} \, \sigma_T \over 2 \, c}
\int\limits_{\mu_2^{\rm min}}^{\mu_2^{\rm max}} d \mu_2
\int\limits_{x_2^{\rm min}}^{x_2^{\rm max}} d x_2 \; {\dot N_D
(\epsilon_1, \Omega_1) \over r^2},
\end{equation}
where $\sigma_T$ is the Thomson cross section, $\mu_2 = \cos\alpha_2$
(cf. Fig. 1), $\epsilon_1 = \epsilon'_2 / D_2$, and $D_2 = \Gamma
(1 + \beta_{\Gamma} \mu_2)$. The term $\dot N_D (\epsilon_1, \Omega_1)$
is the angle-dependent photon production rate of the disk in the
direction $\Omega_1 = (\theta_1, \phi_1)$.  In the derivation, note 
that the photon travels opposite to the direction for which 
$\alpha_2$ is defined. The integration limits $\mu_2^{\rm min/max}$ 
and $x_2^{\rm min/max}$ in equation (1) follow from the geometrical 
constraint that $r_{\rm in} \le r \le r_{\rm out}$, and from the 
requirement $x_s > R'_B$, where $x_s = x_2 \, D_2$ is the distance 
of the location of the scattering event in the BLR to the center 
of the blob  at the time of scattering in the BLR. The latter 
constraint  represents the condition that there is no cold BLR 
material within the relativistic blob.

The synchrotron reflection mechanism (GM\markcite{gm96}96),
where the jet synchrotron radiation is scattered within the BLR and
reenters the jet at a later time  and at a different location $z_2$,
is more complicated because light travel-time effects
have to be carefully taken into account. The comoving-frame 
RSy photon number density at height 
$z_2$ is given by the expression

\begin{equation} {n'}_{\rm ph}^{\rm RSy} (\epsilon'_2; z_2) =
{n_e^{\rm BLR} \, \sigma_T \over 2 \, c}
\int\limits_{\mu_2^{\rm min}}^{\mu_2^{\rm max}} d \mu_2
\int\limits_{x_2^{\rm min}}^{x_2^{\rm max}} d x_2 \> D_1^2 \,
{\dot N'_{sy} (\epsilon'_1, \Omega'_1; z_1) \over x_1^2},
\end{equation}
where $\epsilon'_1 = \epsilon'_2/(D_1 \, D_2)$, $D_1 = \bigl[ \Gamma
(1 - \beta_{\Gamma} \mu_1) \bigr]^{-1}$, and $\mu_1 = \cos\alpha_1$.
$\dot N'_{sy} (\epsilon'_1, \Omega'_1; z_1)$ is the angle-dependent
synchrotron photon production rate in the comoving frame at height
$z_1$, where $\Omega'_1 = (\alpha'_1, \phi'_1)$. In the derivation
giving equation (2), we use the transformation relating the differential
photon production rate between the stationary receiving and comoving frames, 
given by $\dot N_{\rm ph}  (\epsilon_1,
\Omega_1) = D_1^2 \dot N'_{\rm ph} (\epsilon'_1, \Omega'_1)$.

The time of photon emission at location $z_1$ is given by the causality
condition $z_1 = z_2 - \beta_{\Gamma} (x_1 + x_2)$. Therefore, in
addition to  the geometrical constraints described in the previous
paragraph, the condition $z_1 > 0$ has to be met (note that the light
travel-time constraint has been ignored in GM\markcite{gm96}96). In 
particular, causality implies that the photon density due to the 
RSy process is

\begin{equation} {n'}_{\rm ph}^{\rm RSy} = 0 \;,\; \; \; {\rm provided}
\;\;\; z_2 < z_{\rm cr} \equiv r_{\rm in} \, {2 \beta_{\Gamma} \over 1 +
\beta_{\Gamma}} \approx r_{\rm in} \left(1 - {1 \over 4 \Gamma^2}
\right)\;.  \end{equation}
For $z_{\rm cr} < z_2 < r_{\rm in}$, only radiation scattered from within 
a very small angle $\alpha_1^{\rm max} \approx \Gamma^{-1}\, \sqrt{(z_2 
/ \beta_{\Gamma} \, r_{\rm in}) - 1}$ contributes to the rescattered 
photon density in the blob.\footnote[1]{\rm The small solid angle 
illuminated by the RSy mechanism in comparison with the large solid 
angle subtended by the BLR clouds in the ECC process could explain why no 
varying
Ly$\alpha$ emission  correlated with the highly variable UV continuum was 
detected
from 3C 279  (Koratkar et al.\
\markcite{kpup98}1998).} Furthermore, since photons emitted at 
$z = 0$ are at most a distance $\Delta z = c t (1 - \beta_{\Gamma}) 
\approx z / (2 \, \Gamma^2)$ ahead of the blob when it reaches $z$, 
only a fraction $\Delta z / \Delta r_{\rm BLR}$ of the radial 
Thomson depth of an extended cloud can scatter the synchrotron 
radiation from the blob. As viewed in the comoving frame, this means that unless 
$r_{\rm out}\gg \Gamma R^\prime_B $, no extended photon halo (which contributes 
to 
the reflected radiation) will surround the blob as it passes through the 
BLR.

A relativistically moving blob will receive a considerable flux of
scattered synchrotron radiation only when it is passing through the
BLR, i.e. for $r_{\rm in} \lesssim z \lesssim r_{\rm out}$. Assuming that
$\Delta r_{\rm BLR} \gg \Gamma R^\prime_B$, we find that the rescattered
synchrotron photon energy density can be estimated as

\begin{equation} u'_{\rm RSy} \approx 4 \, \Gamma^3 \, \tau_{\rm T,BLR} \,
{R'_B \over \Delta r_{\rm BLR}} \, \left( 1 - {2 \,\Gamma \, R'_B
\over z} \right) \> u'_{\rm sy}.
\end{equation}
Two powers of $\Gamma$ come from Thomson scattering, and a third from 
length contraction and consequent density enhancement of scattering 
material as viewed in the comoving frame.

In addition to the comoving-frame photon energy densities,
$u'_{\rm ph} = m_e c^2 \int_0^{\infty} d\epsilon' \> \epsilon'
\, n'_{\rm ph} (\epsilon')$, we compute the total radiative
power resulting from Compton scattering of the respective photon
fields by calculating the total energy-loss rate of electrons
due to Compton scattering. The decline of the Compton scattering
cross section in the Klein-Nishina  regime, $\epsilon'\gamma
\gtrsim 1$, is roughly accounted for by neglecting the energy-loss of
electrons with Lorentz factor $\gamma$ which scatter
photons of energy $\epsilon' > 1/\gamma$.  The Compton power is

\begin{equation} P'_{rad} = V^\prime_B \, {4 \over 3} \, c \, \sigma_T \, m_e
c^2
\int\limits_{\gamma_1}^{\gamma_2} d\gamma \> \gamma^2 \, n_e (\gamma)
\int\limits_0^{1/\gamma} d\epsilon' \> \epsilon' \, n'_{\rm ph}  (\epsilon'),
\end{equation}
where $V^\prime_B$ denotes the blob volume. It is important to note that
the different Doppler-factor dependences of the observed fluxes from different
processes (especially the SSC and EC scattering processes; see Dermer
\markcite{dermer95}1995) are properly taken into account.

\section{Numerical Results and Discussion}

Photon energy densities and radiative powers due to different processes 
were numerically calculated as a function of the distance of the blob 
from the accretion disk for a wide range of parameters. Figure 2 shows 
an example of our series of simulations, using parameters representative 
of a flat-spectrum radio quasar (FSRQ) and a BL Lac object (BL). Here we let 
$p = 3$ (FSRQ; solid curves) and $p = 2.7$ (BL; shaded curves). This choice 
yields
fairly flat $\nu F_{\nu}$  spectra which can be compared with the peaks of the
broadband $\nu F_{\nu}$  spectral energy distributions  of blazars. The electron
density and blob radius are chosen so that the resulting total luminosities are 
in
accord  with typical values of the apparent luminosities of FSRQs and BLs, and 
so
that 
$< 100$ GeV $\gamma$ rays are not absorbed by 
$\gamma$-$\gamma$ pair production on the synchrotron photons intrinsic 
to the source. 

For the generic FSRQ, we assume that the BLR has a radial Thomson depth
$\tau_{\rm T, BLR} = 0.2$ and occupies a spherical shell located between 
0.05 and 0.5~pc from the central engine. For simplicity, we assume that 
the accretion disk radiates isotropically with a luminosity of 
$10^{46}$~erg~s$^{-1}$. This choice of parameters gives a Compton 
power which is $\sim 10$ times greater than the synchrotron power.  
Figure 2 shows that although the RSy radiation 
energy density $u'_{RSy}$ is larger than $u'_{Sy}$ when the blob 
is located within the BLR, the RSy radiative power is about equal to the SSC 
power.
This is because the backscattered jet synchrotron photons are boosted in energy 
by a
factor of $\Gamma^2$ relative to the energy of the synchrotron photons in the
comoving frame; thus Klein-Nishina effects reduce the radiative power resulting
from Compton scattering of RSy radiation
more strongly than for the SSC radiation.

In our simulations using plausible parameters compatible with the 
assumed spherical shell geometry of the BLR, we find that the Compton power 
due to the synchrotron mirror mechanism is at most comparable to the 
SSC power.  The efficiency of this process is improved when $\Gamma
\gg 10$ and a very narrow ($\Delta r_{\rm BLR}  \lesssim r_{\rm in} 
/ (2\Gamma^2)$), Thomson-thick BLR is located very far ($r_{\rm in} 
\gg \Gamma^2 R'_B$) from the central engine. The synchrotron reflection 
process might therefore operate in BLR clouds which are thought to surround 
Seyfert AGNs and FSRQs. BLR clouds, as understood through photoionization 
models (see, e.g., Wandel \markcite{Wandel97}1997 and references therein), 
consist of dense ($n\sim 
10^{10}-10^{11}$ cm), Thomson-thick ($\tau \sim 1-10$) regions covering a small 
($\sim 10$\%) 
fraction of the central engine. For such a model to be feasible, however, the 
conditions  regarding duty cycle and power outlined by Dermer \& Chiang 
(\markcite{dc98}1998) must be met.  We defer presentation of 
results in the regime $R^\prime_B \gtrsim \Delta r_{\rm BLR}/\Gamma$ 
to future work; the shell geometry used here is excessively 
artificial in this limit. 

The absence of strong emission lines in X-ray selected BLs
and (to a lesser extent) radio-selected BLs suggests that 
the BLR is considerably more dilute in BLs than in FSRQs, and that BLs have mean 
Thomson 
thicknesses $\tau_{\rm T,BLR} \ll 0.1$ (see Scarpa \& Falomo 
\markcite{sf97}1997 and references therein). (On the other hand, 
the strength of the central ionizing photon source might be much 
less in BLs than FSRQs.)  Superluminal motion observations also 
indicate that typical values of $\Gamma$ for BLs lie in the range between $\sim 
3$ 
and 7 (see the review by Urry \& Padovani \markcite{up95}1995).  
The ability of the synchrotron reflection process to produce 
gamma-ray flares therefore seems more difficult in BLs than in FSRQs,
yet TeV flares often coincident with X-ray flares have been detected from three
BL Lac objects (e.g., Punch et al. \markcite{punch92}1992, Macomb et al. 
\markcite{Macomb95}1995; 
Catanese et al. \markcite{Catanese97}1997, \markcite{Catanese98}1998). 
For the assumed BL parameters in Figure 2, the synchrotron 
reflection flare could hardly be detected.

If an accretion disk steadily radiates photons (see B\"ottcher \& Dermer
\markcite{bd95}1995 for a treatment of time-variable disk radiation),
then light-travel time effects can be neglected. The accretion disk supplies an
abundant supply of soft photons, which can enter the jet directly (ECD) and 
after
being scattered by the BLR (ECC). The comoving photon energy  densities from the 
ECD
process can dominate that from the ECC and synchrotron processes when $z 
\lesssim
10^{-2}$~pc, but declines 
$\propto z^{-3}$ and $\propto z^{-2}$ farther out (see Dermer \& Schlickeiser 
\markcite{ds93}1993;  B\"ottcher, Mause \& Schlickeiser
\markcite{bms97}1997). The ECC photon energy density increases
slowly with $z$ when $z < r_{\rm in}$, and begins to decrease for 
$r_{\rm in} \lesssim z \lesssim r_{\rm out}$. Outside the BLR, 
when $z\gtrsim r_{\rm out}$, the energy density asymptotically 
approaches the limiting behavior  $u^\prime_{\rm ECC} \propto z^{-2}$.

When $z \lesssim r_{\rm in}$, the ECC process dominates
over the SSC process provided that

\begin{equation} {\tau_{\rm T, BLR} \over r_{\rm in}^2} \gtrsim {2 \over 3}
{c \, B^2 \tau_B \over L_D \, \Gamma^2} \> \big( {p - 1 \over 3 - p}\big) \>
{\gamma_2^{3 - p} - \gamma_1^{3 - p} \over \gamma_1^{1 - p} -
\gamma_2^{1 - p}},
\end{equation}
where $\tau_B = R^\prime_B \, n_{\rm e,jet} \, \sigma_T$ is the Thomson
depth of the blob. Using the parameters adopted in Figure 2 but letting
$\tau_{\rm T, BLR}$, $r_{\rm in}$, and $L$ vary, we find that the ECC photon 
energy density dominates the synchrotron photon energy density in the 
comoving blob frame when $ r_{\rm in}({\rm pc}) \lesssim 0.4 \>
\tau^{1/2}_{\rm T, BLR} \, L_{46}^{1/2}$ and $ r_{\rm in}({\rm pc}) 
\lesssim 0.1 \> \tau^{1/2}_{\rm T, BLR} \, L_{44}^{1/2}$ for the FSRQ 
and BL parameters, respectively, where $L_n = L_{\rm disk} / (10^n~{\rm
erg \> s}^{-1})$. It should be noted that Klein-Nishina effects are not 
included in estimate (6).

The bottom panel of Figure 2 illustrates the time profiles of a flare  
calculated
using the FSRQ parameters for the ECC (thick solid curve)  and RSy (thick 
dot-dashed
curve) processes.  Note that the observer's time element is linearly related to 
$z$
for  a blob moving with constant velocity. The ECC process gives a fast-rise, 
power-law-decay--type light curve, and the RSy mechanism gives a  gradual rise 
of
the $\gamma$-ray flux and a sharp drop as the blob leaves the BLR. A flare 
produced
by the RSy process could be identified by  a rapid decline of $\gamma$-rays 
which is
not accompanied by a corresponding decrease of the  synchrotron emission. For
comparison, we also sketch a flare time  profile produced by the ECD process,  
and
time profiles produced by  sweeping energization of the blob.  In this process, 
the
bulk kinetic  energy of  the outflowing plasmoid is converted into internal 
nonthermal particle energy by sweeping up BLR material  (see, e.g., Panaitescu 
\&
M\'esz\'aros \markcite{pm98}1998; Dermer \& Chiang
\markcite{dc98}1998;  Chiang \& Dermer \markcite{Chiang98}1998).  The pair of 
light
solid and dot-dashed curves illustrate ECC and RSy flares, respectively, are 
modeled 
assuming that the nonthermal  electron energy is proportional to the amount of
swept-up matter which is then added to a nonthermal lepton distribution 
accelerated at
the base of the jet. Blob  deceleration is assumed to be negligible here.  The 
upper
curves of the two pairs are modeled assuming no radiative losses, and the lower
curves of the two pairs illustrate the effects of radiative losses on the 
nonthermal
leptons by crudely multiplying the upper curves by a decaying exponential with a 
$1.5\times 10^{18}$~cm decay length.

The model RSy light curves, with and without sweeping energization,  are similar 
to
$\gamma$-ray light curves observed in the 1991 and 1996 flares of 3C 279 
(Hartman et
al. \markcite{Hartman96}1996; Wehrle et al. \markcite{Wehrle98}1998). This may 
be a
consequence, however, of the highly idealized BLR  geometry used in the 
calculation.
More symmetical flaring profiles observed from PKS 0528+134 (Collmar et al. 
\markcite{collmar97}1997), PKS 1622-297 (Mattox et al. \markcite{Mattox97}1997), 
and
PKS 1406-076 (Wagner et al.
\markcite{Wagner95}1995) might be more easily explained by the ECD or ECC
processes.  The declines of the X-ray and optical fluxes correlated with the 
EGRET
$\gamma$-ray fluxes in the February 1996 3C 279 and the 1406-076 flares could,
however, rule out the RSy mechanism since the synchrotron component is not 
directly
affected by the reflection process. The inclusion of electron energy evolution 
and the
relaxation of the assumption of a constant velocity blob must be treated to
strengthen such conclusions.  Future flare modeling must treat the  passage of 
the
jet through the BLR clouds, which themselves are in  Keplerian motion around the
central black hole. The passage of a  jet through such a region will display a
complicated signature when monitored by  different telescopes with different
sensitivities  and imaging capabilities, which can only be decoded  when full 
account
is taken of the processes considered here.  The efficiency of the different
Compton-scattering scenarios,  including the RSy mechanism, in this more  
realistic
system is presently under investigation by the authors.  

\section{Summary}

We have analyzed different mechanisms for producing $\gamma$-ray
flares which involve relativistic leptons in blazar jets which 
Compton-scatter soft photons. We presented a detailed study of 
the jet synchrotron reflection process (Ghisellini \& Madau 
\markcite{gm96}1996), where the beamed synchrotron radiation 
emitted by a relativistic plasma blob is scattered back into 
the jet by BLR material, and determined conditions under which 
this mechanism can operate efficiently compared to competing 
mechanisms. Our calculations show that the efficiency of the synchrotron 
reflection mechanism is improved for very narrow ($\Delta r_{\rm BLR} \lesssim 
r_{\rm in} 
/ 2\Gamma^2$) Thomson-thick $\tau_{\rm BLR} \gtrsim 0.1$ scattering regions 
located far ($r_{\rm in}
\gtrsim 0.1 $~pc) from the central accretion disk. We argued that the 
synchrotron reflection
model may have serious difficulties  explaining $\gamma$-ray flares from 
BL-Lac objects.  Model flare time profiles for various 
Compton-scattering processes, which in addition include 
effects of energization of the blob by sweeping up circumnuclear 
material, were illustrated. The slow-rise, rapid-decay time
profile seen in several EGRET flares from FSRQs could be 
realized by the RSy process, though an accompanying flare from 
the ECC and ECD processes would also be seen unless the blob only 
becomes energized after it enters the BLR.  A more detailed 
study of flare production which includes the reflected synchrotron
mechanism and sweeping energization in a system where the BLR 
consists of small, Thomson thick clouds is an important subject 
for further study.

\acknowledgements MB acknowledges support by the German Academic Exchange
Service (DAAD) and by NASA grant NAG~5-4055. The work of CD is supported by
the Office of Naval Research and the {\it Compton Gamma Ray Observatory}
Guest Investigator Program.

\eject

\begin{figure}
\epsfysize=11cm
\rotate[r]{
\epsffile[200 50 550 500]{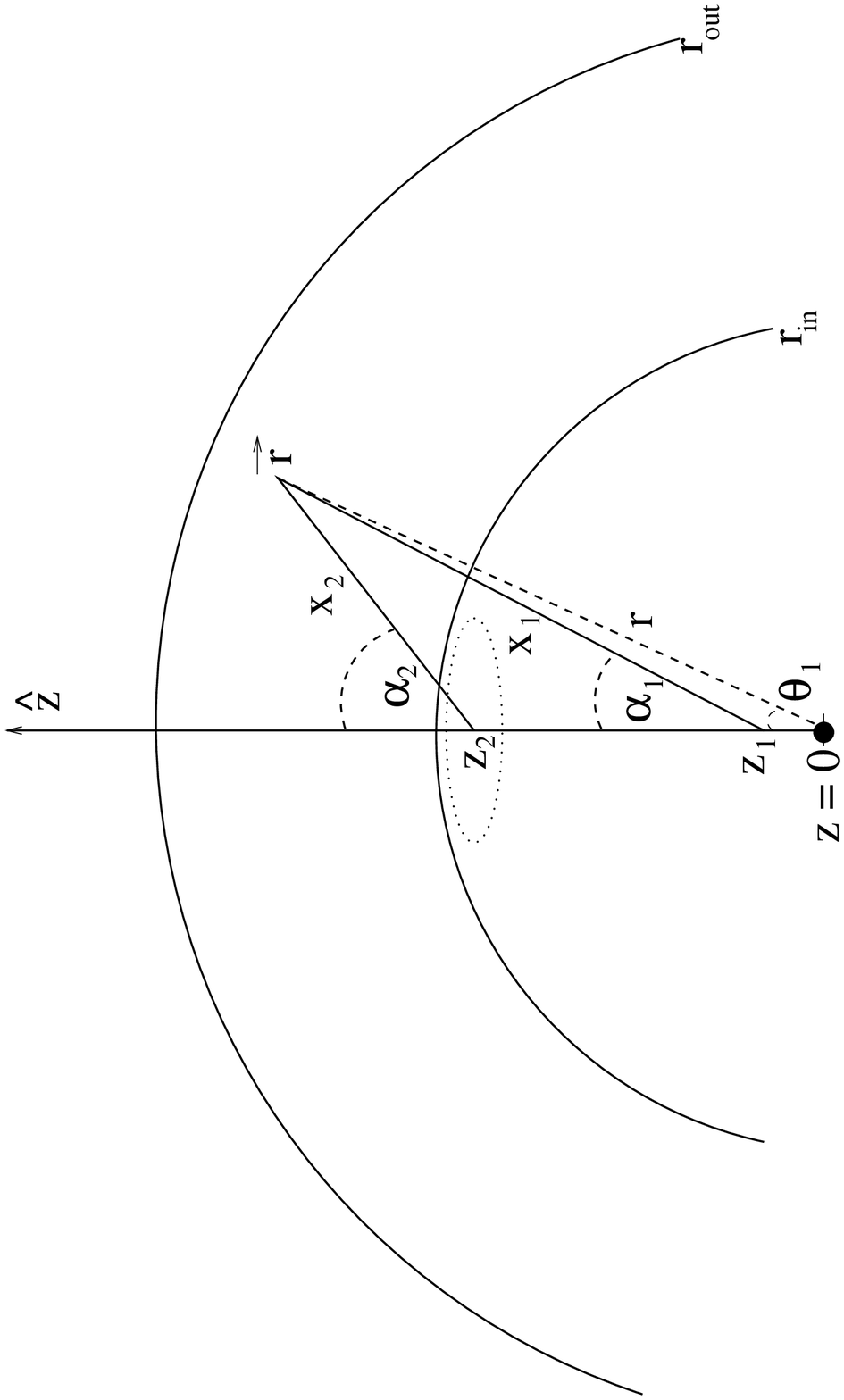}}
\caption[]{Geometry used for the computation of jet radiation rescattered
into the jet trajectory. Radiation is emitted at distance
$z_1$ from the accretion disk at $z = 0$, is rescattered in the BLR at a
distance $r$ from the central engine, and reenters the jet at
location
$z_2$.}
\end{figure}

\eject

\begin{figure}
\epsfysize=12.5cm
\rotate[r]{
\epsffile[150 0 580 550]{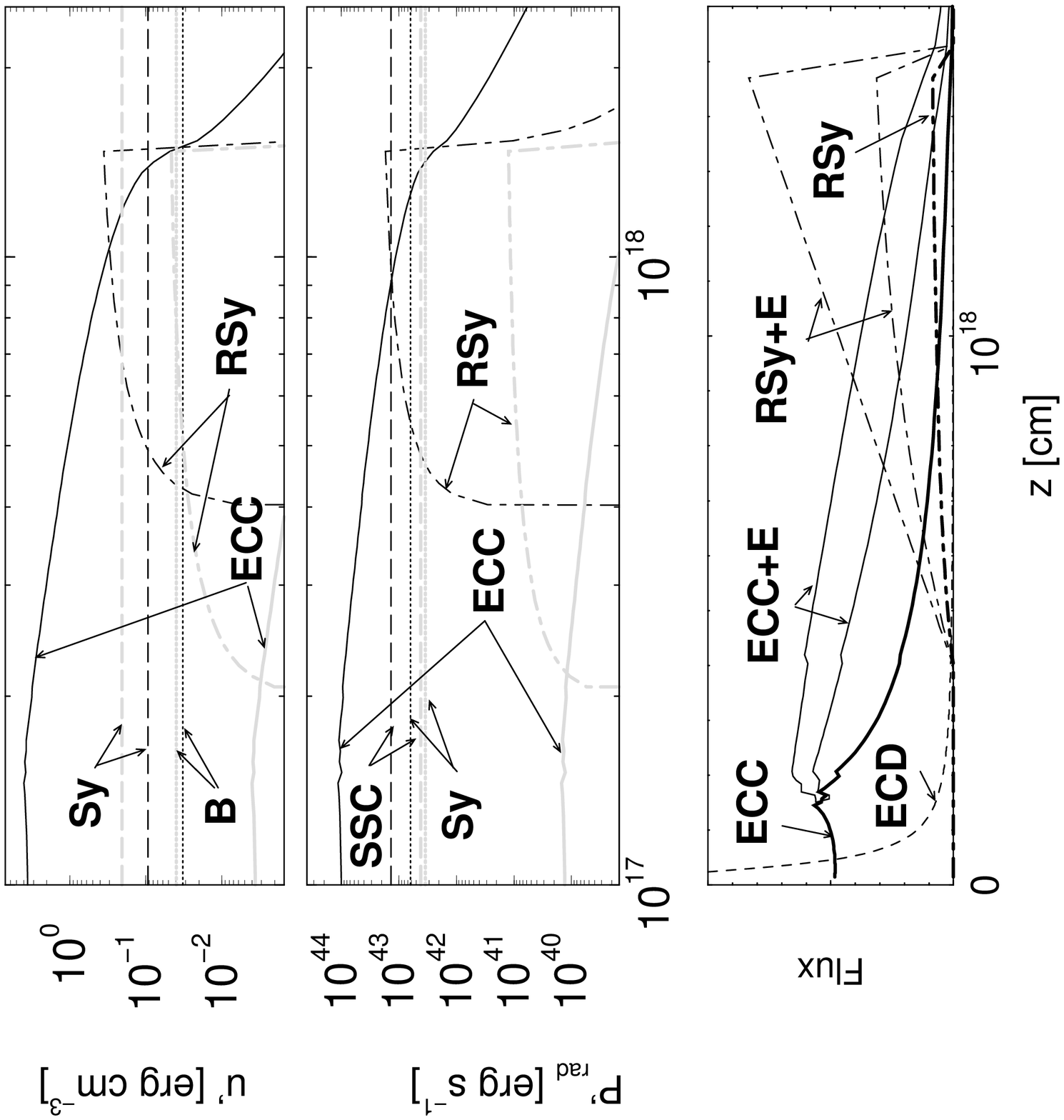}}
\caption[]{Energy densities (upper panel), radiative powers (middle power), 
and model light cuves (bottom panel) for synchrotron and Compton processes, 
plotted as a function of distance $z$ of the blob from the central engine. 
`ECC' denotes accretion-disk photons scattered from the BLR into the blob, 
`RSy' denotes reflected synchrotron photons, `ECD' denotes accretion-disk 
photons entering the blob directly, and `+E' curves illustrate the effects 
of sweeping energization on the blob as it passes through the BLR. Standard 
parameters used for a typical FSRQ (solid curves) are $\Gamma = 10$, 
$\gamma_1 = 10^3$, $\gamma_2 = 2 \cdot 10^5$, $p = 3.0$, $n_{\rm e,jet} = 
20$~cm$^{-3}$, $B = 0.91$~G, $R'_B = 2 \cdot 10^{16}$~cm, 
$r_{\rm in} = 0.05$~pc, $r_{\rm out} = 0.5$~pc, $\tau_{\rm T,BLR} = 0.2$, 
and $L_{\rm disk} = 10^{46}$~erg~s$^{-1}$. 
Standard parameters used for a typical BL (shaded curves) are the same as 
for the FSRQ case except that $\gamma_2 = 10^6$, $p = 2.7$, $R'_B = 
10^{16}$~cm, $B =  0.99$~G, $\tau_{\rm T,BLR} = 0.02$, and 
$L_{\rm disk} = 10^{44}$~erg~s$^{-1}$. 
Bottom panel uses a linear scale and FSRQ parameters.}
\end{figure}


\begin{references}

\reference{bk79} Blandford, R. D., \& K\"onigl, A. 1979, ApJ, 232, 34

\reference{bl95} Blandford, R. D., \& Levinson, A. 1995, ApJ, 441, 79

\reference{bm96} Bloom, S. D., \& Marscher, A. P., 1996, ApJ, 461, 657

\reference{bloom97} Bloom, S. D., Thompson, D. J., Hartman, R. C., \& von
Montigny, C., 1997, in 4th Compton Symposium,  ed. C. D. Dermer, M. S.
Strickman, \& J. D. Kurfess (New York: AIP), 1262

\reference{bd95} B\"ottcher, M., \& Dermer, C. D., 1995, A\&A 302, 37

\reference{bms97} B\"ottcher, M., Mause, H., \& Schlickeiser, R., A\&A, 324, 395

\reference{Catanese97} Catanese, M., et al. 1997, ApJ, 487, L143

\reference{Catanese98} Catanese, M., et al. 1998, ApJ, in press

\reference{Chiang98} Chiang, J., \& Dermer, C. D. 1998, ApJ, submitted

\reference{collmar97} Collmar, W., et al.,
1997, in Proc. of the XXV.\ ICRC, Vol. 3, 101, ed. M. S. Potgier, 
B. C. Raubenheimer, \& D. J. van der Walt

\reference{dermer95} Dermer, C. D., 1995, ApJ, 446, L63

\reference{dc98} Dermer, C. D., \& Chiang, J., 1998, New Astronomy,  3, 157


\reference{ds93} Dermer, C. D., \& Schlickeiser, R., 1993, ApJ, 416, 458


\reference{dss97} Dermer, C. D., Sturner, S. J., \& Schlickeiser, R., ApJS, 109,
103

\reference{gm96} Ghisellini, G., \& Madau, P., 1996, MNRAS 280, 67

\reference{Hartmen96} Hartman, R. C., et al. 1996, ApJ, 461, 698

\reference{Hartman97} Hartman, R. C., Collmar, W., von Montigny, C.,
\& Dermer, C. D., in The Fourth Compton Symposium, ed. C. D. Dermer, M. S.
Strickman, \& J. D. Kurfess (New York: AIP), 307

\reference{hummel97} Hummel, C. A., Krichbaum, T. P., Witzel, A., et al., 1997,
A\&A, 324, 857

\reference{kpup98} Koratkar, A., Pian, E., Urry, C. M., \& Pesce, J. E. 1998,
ApJ, 492, 173

\reference{Macomb95} Macomb, D. J. et al. 1995, ApJ, 449, L99; (e) 1996, 459, 
L111

\reference{maraschi92} Maraschi, L., Ghisellini, G., \& Celotti, A., 1992, ApJ,
397, L5

\reference{mg85} Marscher, A. P., \& Gear, W. K., 1985, ApJ, 298, 114

\reference{Marscher90} Marscher, A. P. 1990, in Parsec-Scale Radio Jets, ed. J.
A. Zensus \& T. J. Pearson (New York: Cambridge University Press), 236

\reference{Mattox97} Mattox, J. R., et al., 1997, ApJ, 481, 95

\reference{mukherjee97} Mukherjee, R., Bertsch, D. L.,  Bloom, S. D.,
et al., 1997, ApJ 490, 116

\reference{pm98} Panaitescu, A., \& M\'esz\'aros, P. 1998, ApJ, 492, 683

\reference{pb97} Protheroe, R. J., \& Biermann, P. L. 1997, Astroparticle 
Physics, 6, 293

\reference{punch92} Punch, M., et al., 1992, Nature 358, 477


\reference{sf97} Scarpa, R., \& Falomo, R. 1997, A\&A, 325, 109

\reference{schlickeiser96} Schlickeiser, R., 1996, Space Sci. Rev., 75, 299

\reference{ss73} Shakura, N. I., Sunyaev, R. A., 1973, A\&A, 24, 337

\reference{sbr94} Sikora, M., Begelman, M. C., \& Rees, M. J., 1994,  ApJ, 421,
153
\reference{steffen95} Steffen, W., et al., A\&A,
302, 335

\reference{up95} Urry, C. M., \& Padovani, P. 1995, PASP, 107, 803

\reference{wagner95} Wagner, S. J., et al. 1995, ApJ, 454, L97

\reference{Wandel97} Wandel, A. 1997, ApJ, 490, L131

\reference{Wehrle98} Wehrle, A., et al.\ 1998, ApJ, in press

\end{references}
\end{document}